# Miniaturized metalens based optical tweezers on liquid crystal droplets for lab-on-a-chip optical motors

Satayu Suwannasopon,[1,2,3] Fabian Meyer,[3] Christian Schlickriede,[3] Papichaya Chaisakul,[1] Jiraroj T-Thienprasert,[1] Thomas Zentgraf,[3] and Nattaporn Chattham[1,2]

[1] Department of Physics, Faculty of Science, Kasetsart University, Bangkok 10900, Thailand

[2] Center for Advanced Studies in Nanotechnology and its Applications in Chemical, Food, and Agricultural Industries, Kasetsart University, Bangkok 10900, Thailand

[3] Department of Physics, Paderborn University, D-33098 Paderborn, Germany

**Introduction**

Metamaterials are artificial materials with their unusual properties emerged from structural arrangement rather than composition. By directly designing the arrangement of meta-atoms, extraordinary properties of light propagation beyond nature availability can be achieved.[1-3] For over twenty years, this emerging field of metamaterials has capture enormous interest from scientists to explore unique optical effects such as cloaking, negative refraction or ultrathin lenses. Powerful and sophisticate electromagnetic modeling software, nowadays, is able to accurately simulate and predict the response of pattern design before fabrication attempt. The success in creating metamaterials in the visible light region was first reported in 2007.[4] A novel type of metamaterial consisting of planar arrays of artificial nanostructures on ultrathin layers – so-called metasurfaces – capable of light manipulation have recently been discovered.[5-8] Most metasurfaces studied so far typically comprise of plasmonic nanoantennas of designed shape and orientation pattern on thin substrates. The optical property difference between the antennas and surrounding medium results in optical interaction generating and controlling the dispersion of light, amplitude, phase profile and polarization response along the interface.[9] Recently, several potential applications of metasurfaces have been suggested, which have been led to the development of ultrathin devices such as waveplates for generating vortex beam,[10,11] beam steering, holography,[6] aberration-free quarter waveplates (QWPs), spin-hall effect of light and spin-controlled photonics,[12] unidirectional surface plasmon polariton excitation[7,13-15] and ultrathin metalenses.[6,16]

Metalenses have emerged from the advance in the development of metasurfaces providing a new basis for recasting traditional lenses into thin, planar optical components, having similar or even better performance at smaller scales. In the near future with higher efficiency and cheaper production,[17] these metalenses have a high potential to replace traditional lenses in several applications, e.g., imaging systems, optical data storage, laser printing, and optical communications. Their ultrathin sizes can also benefit the development of optical Lab-on-a-chip (LOC) devices. The aim of this work is to evaluate the potential of plasmonic metasurfaces acting as a phase mask on a substrate material to modify the beam shape of laser light for polarized optical trapping,[18,19] following the recent success on the first planar metalens optical trap in microfluidic system.[20] For most laboratories, drawbacks of optical trapping have been the high cost involving large equipment like a microscope as well as the difficulty to build and maintain such systems. To overcome these inconveniences, we applied metalenses to create a circularly polarized optical tweezer system as a handheld device and test it on a particle trap and localization of birefringent liquid crystal droplets.

## Materials and method
### Design and fabrication of the metalens

We evaluate the potential of a plasmonic metalens to tightly focus the beam shape of laser light for optical trapping and particle manipulation. Laser light that passes through the metasurface experiences arbitrary continuous phase gradients that can be spatially tailored by utilizing the accumulated Pancharatnam-Berry phase during a polarization conversion for circularly polarized light.[21] For this concept, each nanostructure acts like a local half waveplate, which introduces an orientation-dependent phase to the light. In order to fabricate the structures of the plasmonic metasurfaces, we performed design operation with MATLAB FDTD and CST Microwave Studio. Based on the concept of interfacial phase discontinuities,[8] these software were used for configuring lens pattern and simulation of the scattering response of plasmonic nanoantennas. Simulation results for a metalens pattern of 800 µm focal length are shown in Figure 1. Arrays of plasmonic nanorods made of gold were fabricated on a glass substrate by using electron beam lithography (EBL). For the fabrication process, a thin layer of indium tin oxide (ITO) was used on top of the substrate to obtain conductivity for the EBL. After writing the structure with EBL, the electron beam resist was used as a mask for subsequent thermal evaporation of a 40 nm thick gold layer and which was later removed by a lift-off process. Based on this process, a metalens for 1064 nm left circularly polarized light incident beam was obtained with diameter 300 µm and focal length of f = 800 µm with N.A. = 0.6, as shown in a photomicrograph in Figure 2(a). A scanning electron microscopy (SEM) image of the lens (Figure 2(b)) illustrates each nanorod of dimension 90 x 220 x 40 nm$^3$ placed with a defined orientation angle $\varphi$ serving as meta-atom. In order to function as a conventional spherical lens, the phase profile $\varphi(x, y)$ of the metalens is calculated by the following equation[16]:

$$\varphi(x, y) = \frac{2\pi}{\lambda}(\sqrt{x^2 + y^2 + f^2} - f) \tag{1}$$

where $f$ is the focal length of the lens, $\lambda$ is wavelength in free space, $x$ and $y$ are coordinates of each nanorod. The lens was tested with Fourier Transform Infrared Spectroscopy (FTIR) showing resonance of transmission spectra at 1064 nm (Figure 2(c)) according to the design.

## Optical trapping of particles with the metalens

Optical tweezers are normally formed by tightly focused laser beams with a high numerical aperture objective lens. It has been widely used for manipulation of small particles without direct mechanical interaction with the surrounding. The laser beam can trap particles based on two types of forces. One is the scattering force $\vec{F}_s$, force due to the change in velocity of light that occurs when a laser beam travels between two different mediums explained by the following of equation[22]:

$$\vec{F}_s = n_m \frac{\sigma \langle \vec{S} \rangle}{c} = n_m \frac{\sigma I \langle \vec{r} \rangle}{c} \hat{z} \qquad (2)$$

where $\langle \vec{S} \rangle$ is the time-averaged Poynting vector which is equal to the intensity of incident beam, $\sigma$ is the scattering cross-section of trapped object depending on the ratio of refractive index $(n_p/n_m)$; $n_p$, $n_m$ is the refractive index of particle and medium respectively and $c$ is the speed of light in the vacuum.

The other is the gradient force $\vec{F}_{grad}$, caused by the law of conservation of momentum, resulting from the interaction of photon pressure of the incident laser beam against the surface of the trapped particle. Stable trapping arises from the gradient force by diverging the beam away from the focal point.

$$\vec{F}_{grad}(\vec{r}) = \frac{1}{2n_m \varepsilon_0 c} \alpha \vec{\nabla} I(\vec{r}) \qquad (3)$$

where $I(\vec{r})$ is intensity, $\varepsilon$ is dielectric constant in vacuum and $\alpha$ is the polarizability which depends on the refractive index ratio ($n_p/n_m$).

The experimental setup used for optical trapping of a polystyrene bead (Polysciences, Inc.) with the metalens is shown in Figure 3. A CW 1064 nm fiber laser operated at 300 mW is used as a light source. The incident beam was made circularly polarized by a Glan-Thompson polarizer and a quarter waveplate. The sample chamber is made from a glass slide chamber filled with 4.5 µm polystyrene beads in water and glued on top with metalens (Figure 4). A convex lens is inserted in between a quarter waveplate and the sample to ensure that laser beam covers all surface area of metalens for high efficiency of nanoantenna radiation. The movement of the beads is observed by a CCD camera with a 40X objective lens. The output light was tested to be circularly polarized by inserting another set of a quarter waveplate and a linear polarizer. To find the trapping strength of the optical tweezers, the trapping laser was modulated by frequency modulator in the range of 1 to 10 kHz with a step size of 0.5 kHz[23]. A tracker software was used to track the beads and locate their position in the laser trap at different modulation frequencies.

A circularly polarized laser trap can perform its great advantage on rotating birefringence object[19] by conservation of angular momentum law. The setup was tested with birefringent nematic liquid crystal (NLC) droplets filled inside the metalens chamber. The droplets were prepared by dispersion of 5CB (4-cyano-4'-pentylbiphenyl, SIGMA-ALDRICH) in water with CTAB (Hexadecyltrimethyl ammonium bromide) as a surfactant. By controlling the CTAB concentration, either bipolar or radial NLC droplet or both configurations can be achieved.[24,25]

**RESULTS AND DISCUSSION**

The interaction of the left circularly polarized laser beam with 30 mW power passing through the metalens resulted in the intensity profile of the focal spot at different locations from the lens as shown in Figure 4(a). The laser spot was focused sharply at 800 µm corresponding to the lens design. Interpolation of the intensity measurement at the focal length is depicted in Figure 4(b) with a full width at half maximum (FWHM) spot size of 5.8 µm. The focal spot intensity obtained by a Lumerical FDTD simulation is illustrated in Figure 5 showing the beam profile with FWHM closely similar to one obtained from experimental observation of Figure 4. The optical trapping experiment was then performed with 4.5 µm polystyrene beads in water that were introduced into the sample chamber shown in Figure 3. The trapping experiments were successfully conducted with 30 mW laser power as shown in Figure 6.

We further explored the trapping strength of the laser beam through the metalens by modulation of the laser frequency. The technique reported by Joykutty et al.[23] resembled the phenomenon of parametric resonance in a harmonic oscillator. The measured variance position of the bead in the trap as a function of modulation frequency is plotted in Figure 7 without scaling since the main focus is on relative changes of the value only. The Gaussian fit of the data yields a center frequency of 6.45 ± 0.12 kHz. The trap stiffness of optical tweezers at 30 mW power is calculated from $k_{trap} = 4\pi^2 f_{trap}^2 m$, where $m$ is bead mass of $5\times10^{-14}$ kg, giving the trapping strength $k_{trap}$ = 82 µN/m. With the micron-sized beads used in the experiment, the trapping force of this setup was in the order of a few hundred piconewtons.

To fully employ the circularly polarized focus of the metalens, we then further applied this miniaturized metalens based laser tweezers system on birefringent NLC droplets. NLC droplets suspended in water were filled into the sample chamber. The droplets with radial configuration captured +1 topological defect in the middle while bipolar droplets expelled +1 defect to their surface as illustrated in Figure 8. With our metalens, we observed that trapping of NLC droplets can easily be performed as shown in a video sequence in Figure 9. With the closer observation of the droplet texture under crossed polarizers, we found radial droplets with crossed brushes in the middle and bipolar droplets possessing two boojums on the edge. These two types of droplets behaved differently under the LCP optical trap. Figure 10 shows an image sequence of the radial droplet in the optical trap. The droplet was shaken while in the optical trap with no change observed in the crossed brush texture which is typical for a symmetric configuration under rotation. Thus, with radial droplets, it is difficult to determine whether droplet rotates or not. The observation of bipolar droplets, however, is much easier to determine as the rotation can be observed from the boojum position. Figure 11 shows both radial and bipolar droplets trapped in the laser spot. The bipolar droplet shows clear rotation in the optical trap with the texture changing all the time. With 30 mW laser tweezers, the rotation appeared to be confined in the observation plane. However, by reducing the laser power, the boojums appeared to rotate into the top view and in irregular manner since the force from circularly polarized laser trap no longer dominates. The irregular rotation was influenced by both radiation pressure of the laser and heat convection. The rotation of birefringent droplets is a strong evidence that the optical trap possess strong enough angular momentum of light from radiation of each nanostructure acting like a local half waveplate and introducing an orientation-dependent phase to light. This is the next step of success from the first report of metalens based optical trap[20] by adding the state of polarization of light into the optical trap.

    In summary, we have demonstrated an optical rotation of microdroplets using metalens based polarized optical tweezers on birefringent objects. Ultrathin metalenses are ideal for lab-on-a-chip systems and have a high potential to replace conventional lenses in the near future. Liquid crystal integrated well into this miniaturized optical tweezers device creating an ideal optical motor which can easily be applied into several microscopic systems for motion and flow control.

This project has received funding from the European Research Council (ERC) under the European Union's Horizon 2020 research and innovation programme (grant agreement No 724306). N.C. acknowledges the funding from the Commission on Higher Education, Ministry of Education (the

"National Research University Project of Thailand (NRU)"). N.C. and T.Z. acknowledge the funding from National Research Council of Thailand (NRCT)-Deutsche Forschungsgemeinschaft DFG (ZE953/8-1).

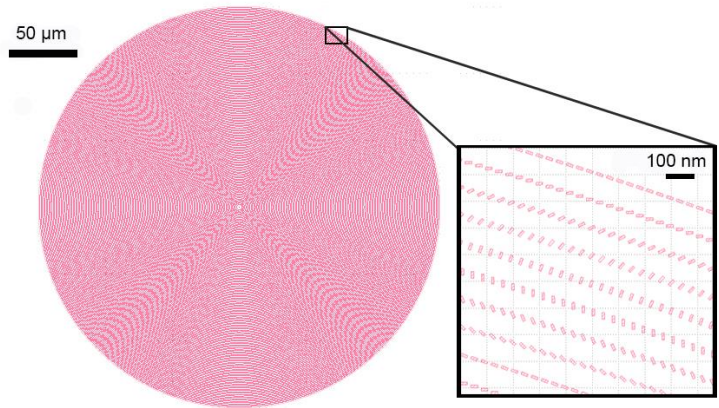

Figure 1 Simulated pattern of 800 µm focal length metalens with a closer view of nanoantenna alignment at the edge of the lens.

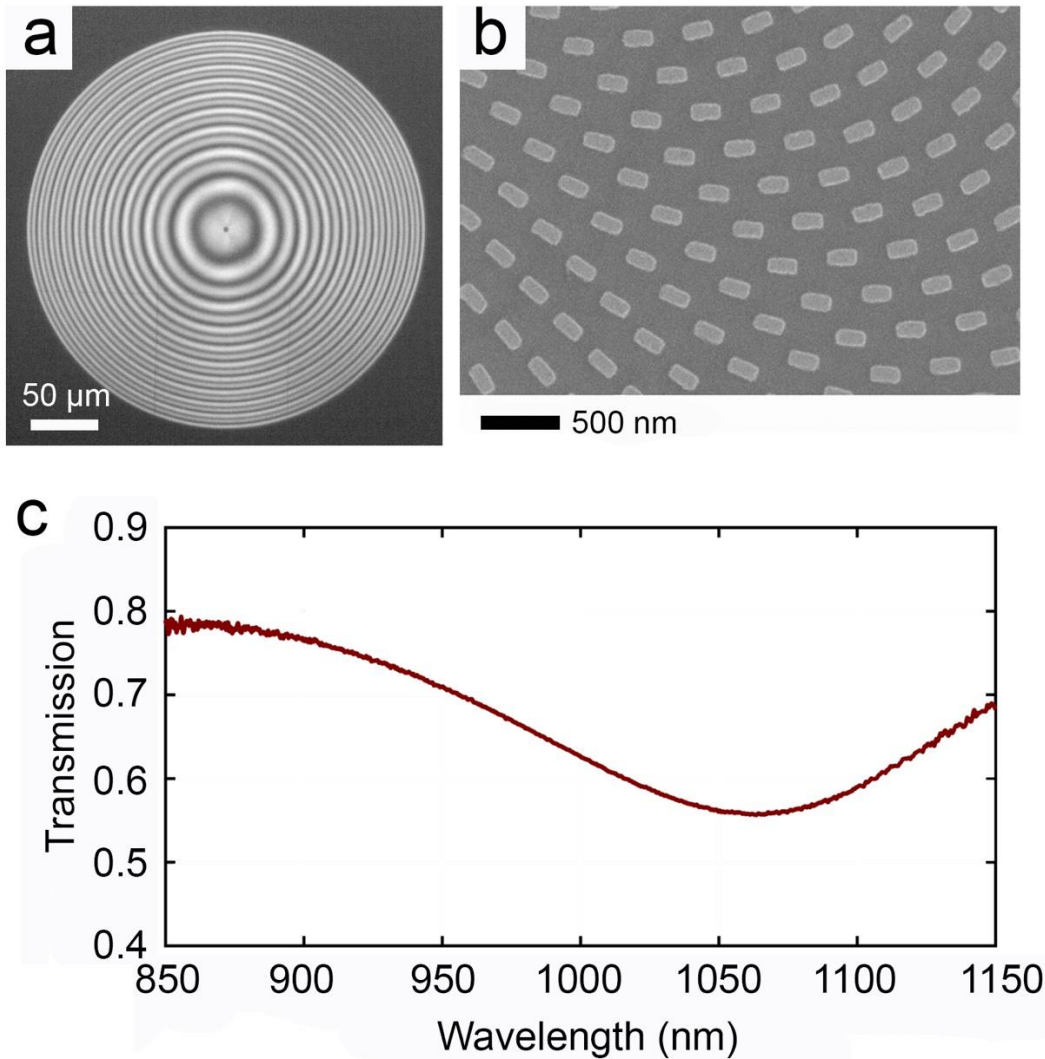

Figure 2 The schematic of plasmonic metalens designed for the focal length of 800 µm. (a) Optical microscopy image of the metalens surface. The scale bar is 50 µm, (b) SEM image of part of the lens showing the gold nanorods. The scale bar is 2 µm. (c) Measurement of the FTIR transmission spectra of the metalens showing a resonance at the wavelength of 1064 nm.

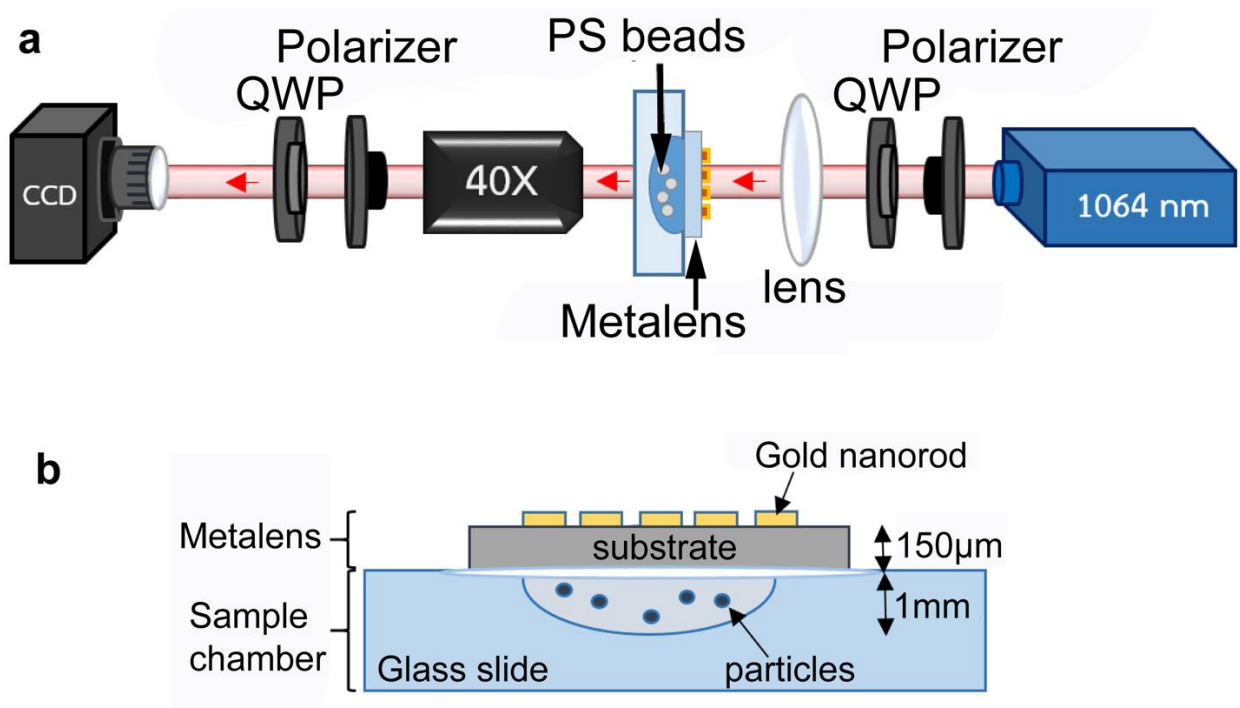

Figure 3 (a) An experimental setup of metalens based optical tweezer. (b) Schematic sketch of the chambered glass slide which was attached on top with metalens as a sample container.

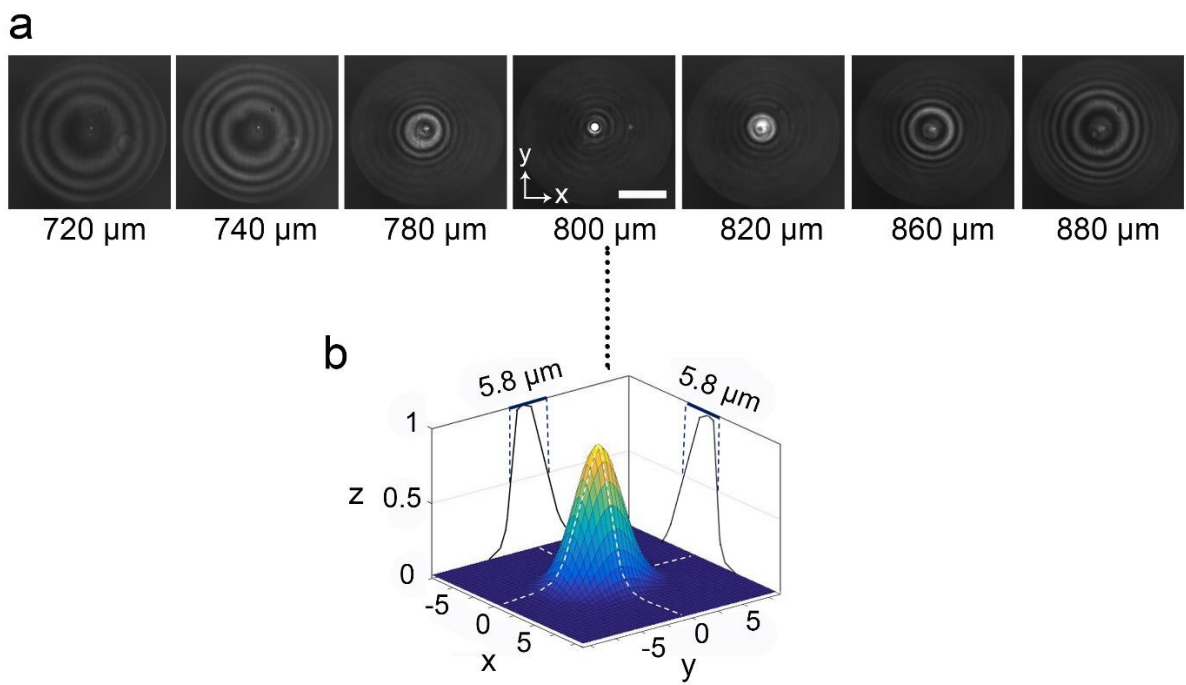

Figure 4 (a) The measurement of the intensity distribution at different locations from the metalens surface for 1064 nm laser light with 30 mW power. The scale bar is 50 µm. The laser beam was focused sharply at 800 µm corresponding to the design. (b) The interpolated surface plot from the intensity measurement at 800 µm gives a full width at half maximum (FWHM) spot size of 5.8 µm.

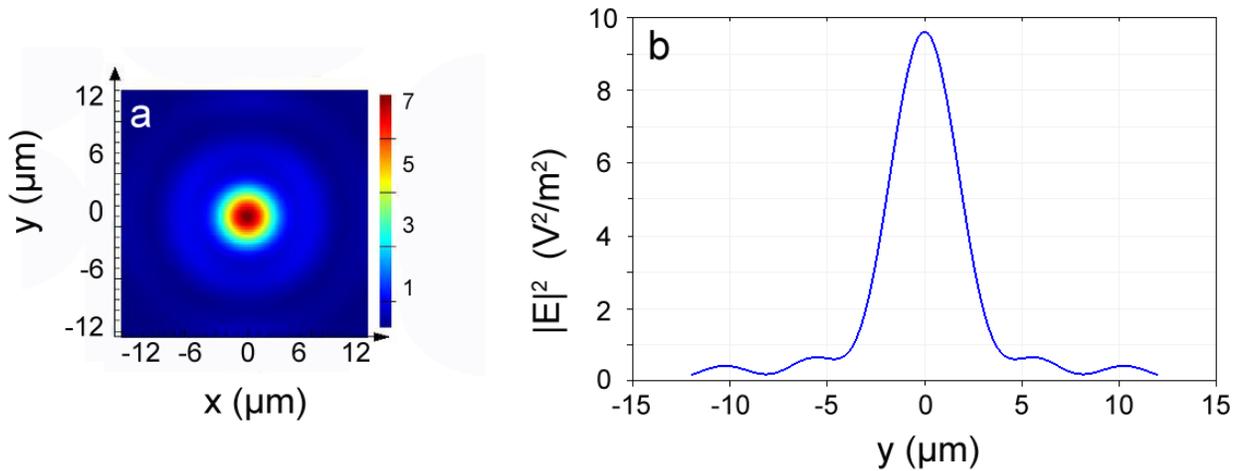

Figure 5 (a) The calculated intensity profile of the focal spot of the metalens calculated by Lumerical FDTD simulation and (b) its vertical cut of the beam through x-z plane with an FWHM of 5.4 µm.

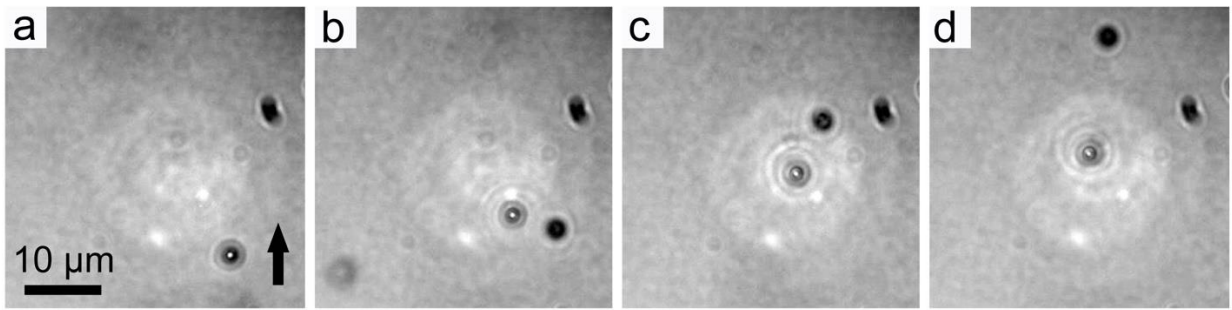

Figure 6  A video sequence recorded during trapping of a PS bead: (a) – (b) bead flowed up due to heat convection of laser spot (c) bead entered into the trap and (d)-(e) remained in the trap while others on the background kept flowing up. Arrow indicates the bead of interest flowing in an upward direction.

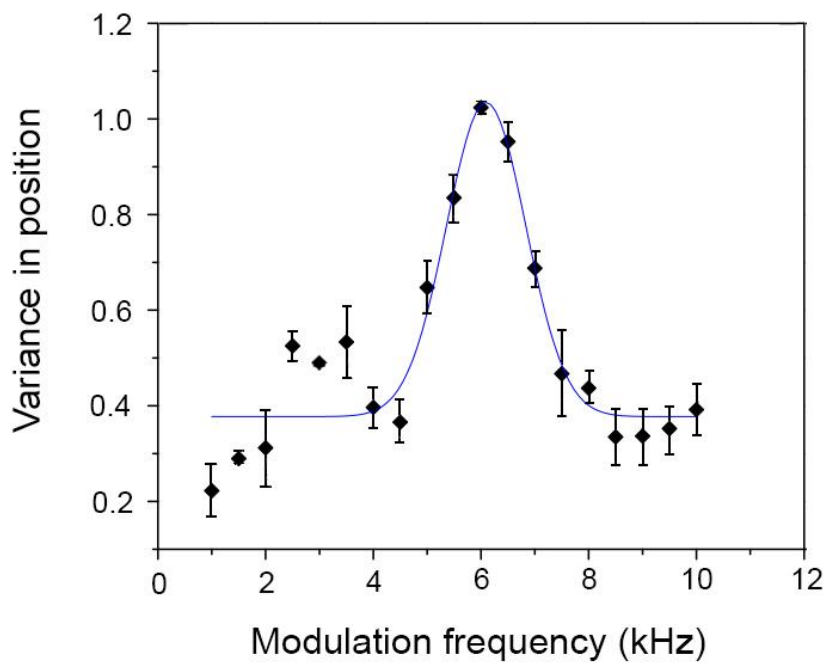

Figure 7 The variance measurement of the bead position as a function of laser modulation frequency.

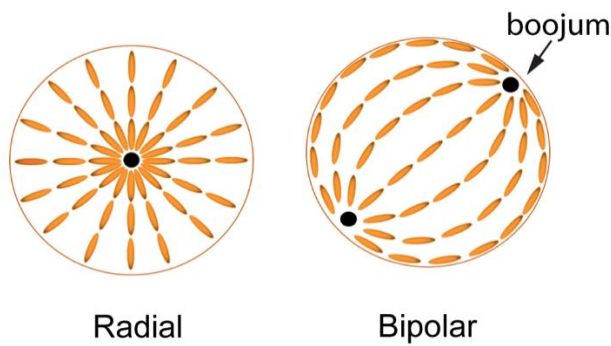

Figure 8 Configurations of radial and bipolar droplets. In the radial droplet, molecules lie perpendicular to the droplet surface trapping a +1 or hedgehog defect in the middle while in the bipolar droplet, molecules lie parallel to the surface confining two boojums at the surface.

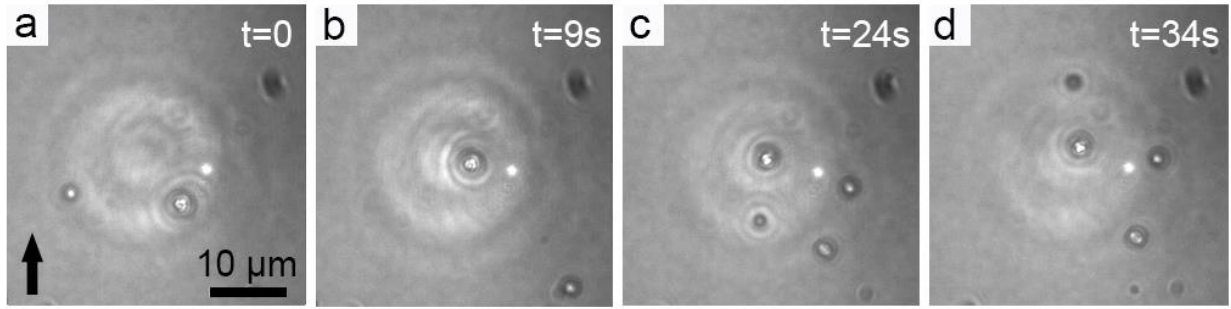

Figure 9 A video sequence showing (a)-(b) an NLC droplet moving upward due to heat convection and was trapped by metalens based laser tweezers. (c)-(d) The trapped droplet remained in the laser focus while others in the background kept flowing upward.

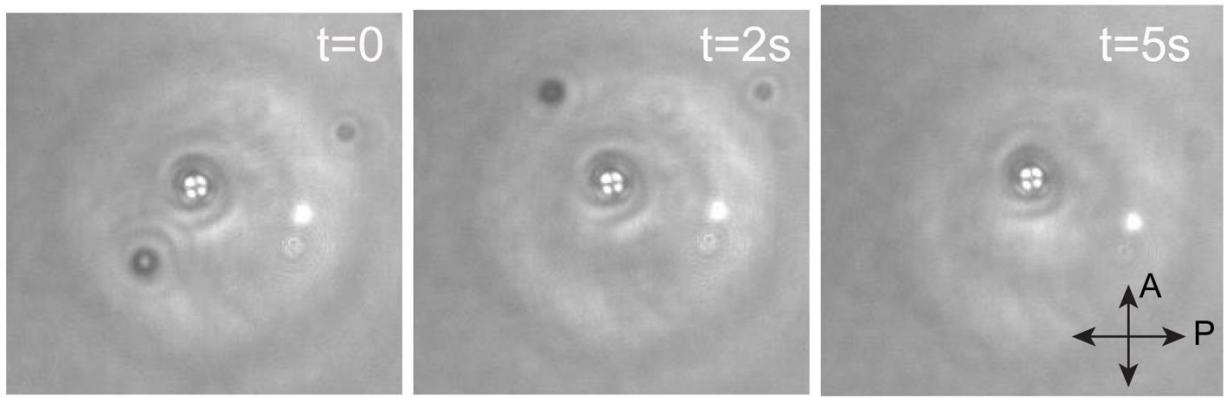

Figure 10 An image sequence of the radial droplet trapped by the optical tweezer. Crossed brushes represent +1 defect confined in the middle of the droplet.

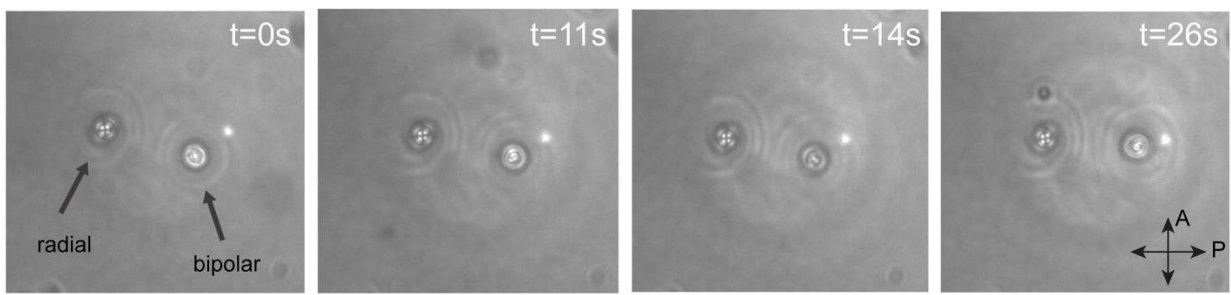

Figure 11 Both radial and bipolar droplets were trapped by the optical tweezers. Radial droplet texture remains the same due to its symmetric configuration while bipolar droplet showed continuous rotation in the laser trap.